\documentclass[12pt]{article}\usepackage{RelaxMethod}

\begin{document}

\title{Relaxation Method\\
For Calculating \\
Quantum Entanglement}

\author{Robert R. Tucci\\
        P.O. Box 226\\ 
        Bedford,  MA   01730\\
        tucci@ar-tiste.com}

\date{ \today} 

\maketitle

\vskip2cm
\section*{Abstract}
In a previous paper, we showed how entanglement of formation
can be defined 
as a minimum of the quantum conditional mutual information 
(a.k.a. quantum conditional information transmission).
In classical information theory, the Arimoto-Blahut method is one of the
preferred methods for calculating extrema of mutual information.
In this paper, we present a new method, akin to the Arimoto-Blahut method, for calculating 
entanglement of formation.  We also present
several examples computed with 
 a computer program called Causa  Com\'{u}n that implements the ideas of this paper.

\newpage
\section{Introduction}

This paper continues a series
of papers\cite{Tucci-tang99}-\cite{Tucci-tang00b} investigating the connection between 
quantum entanglement and conditional information transmission (a.k.a
conditional mutual information, abbreviated CMI). In the last paper of that series,
we expressed the entanglement of formation as a minimum of the quantum CMI. Eureka!
In classical information theory, one of the preferred methods for 
numerically calculating extrema of mutual information is the Arimoto-Blahut algorithm
\cite{Ari}-\cite{Bla-book}.  One wonders whether something akin to that algorithm 
can be used to calculate entanglement. After much huffing and puffing,
we have found the answer to be yes. In this paper we present our first results.
More specifically, we present a new algorithm that yields
 the entanglement of formation
of any bi-partite density matrix 
and a corresponding optimum decomposition of that density matrix. 
Generalization of the algorithm to n-partite systems appears 
straightforward but  we do not address it here.
We also describe a 
C++ computer program called Causa Com\'{u}n that implements the ideas of this 
paper. Finally, we present some examples computed with Causa Com\'{u}n. 

Prior to us, as far as we know, only one group of researchers\cite{Osawa}
has ever used a quantum version of the Arimoto-Blahut algorithm.
They used it to calculate quantum channel capacities.

There exist other excellent computer programs, written prior to ours,  that can 
calculate various features of quantum entanglement. See Ref.\cite{Zyc} and \cite{dutch-boys}.
The software described in Ref.\cite{dutch-boys} also calculates
entanglement of formation and optimal decompositions, but it uses a 
conjugate gradient method that is very different from ours.

\section{Notation}
In this section, we will introduce certain notation which is 
used throughout the paper. 

Let $Bool = \{ 0, 1\}$.
For any finite set $S$, let $|S|$ denote the number of elements
in $S$. 
The 
Kronecker delta function $\delta(x,y)$ equals one if $x=y$ and zero otherwise.
We will often abbreviate $\delta(x,y)$ by $\delta^x_y$. 
Sometimes we will replace an index by the symbol ``$\bullet$". By this we
mean that all values of the index are included.
For example, if we are dealing with $w_\alpha$ where $\alpha\in \{1,2, \ldots n\}$, 
$w_\bullet$ will represent the vector $(w_1, w_2, \ldots, w_n)$.
For any Hilbert space $\hil$,
$dim(\hil)$ will stand for the dimension of $\hil$.
If $\ket{\psi}\in \hil$, then we will often represent 
the projection operator $\ket{\psi}\bra{\psi}$ by $\pi(\psi)$. 

We will underline random variables. For example,
we might write $P(\rvx = x)$ or $P_\rvx(x)$ for the probability that
the random variable $\rvx$ assumes value $x$.
$P(\rvx=x) = P_\rvx(x)$ will often be abbreviated by $P(x)$ when no 
confusion is likely. $S_\rvx$ will denote the
set of values which the random variable $\rvx$
may assume, and $N_\rvx$ will denote the number of
elements in $S_\rvx$. With each random variable $\rvx$,
we will associate an orthonormal basis $\{ \ket{x} | x\in S_\rvx \}$
which we will call the {\it $\rvx$ basis}. 
$\hil_\rvx$ will represent the Hilbert space spanned by the $\rvx$ basis.
For any $|\psi_\rvx\rangle \in \hil_\rvx$,
we will use $\psi_x$ to represent  $\av{x|\psi_\rvx}$.

For any two random variables $\rvx$ and $\rvy$,
$S_{\rvx, \rvy}$ will represent the direct product set 
$S_\rvx \times S_\rvy = \{(x,y) | x\in S_\rvx, y\in S_\rvy\}$. Furthermore, 
$\hil_{\rvx, \rvy}$ will represent $\hil_\rvx \otimes \hil_\rvy$, the tensor product of 
Hilbert spaces
$\hil_\rvx$ and $\hil_\rvy$.
If $\ket{x}$ for all $x$ is the $\rvx$ basis and 
$\ket{y}$ for all $y$ is the $\rvy$ basis, then 
$\hilxy$ is the vector space 
spanned by 
$\{\ket{x,y} | x\in S_\rvx, y\in S_\rvy\}$, where $\ket{x,y} = \ket{x}\ket{y}$.

$\pd(S_\rvx)$ will denote the set of all probability distributions
$P_\rvx$ for the random variable $\rvx$; i.e., 
all functions $P_\rvx:S_\rvx\rarrow[0,1]$ such that $\sum_x P_\rvx(x) = 1$. 
$\dm(\hilx)$ will denote the set of all density matrices $\rho_\rvx$ acting on the  Hilbert space $\hilx$; i.e., 
the set of all $N_\rvx$ dimensional Hermitian matrices with unit trace and non-negative eigenvalues.

Whenever we use the word ``ditto", as in ``X (ditto, Y)", we mean
that the statement is also true if X is replaced by Y. For example,
if we say ``A (ditto, X) is smaller than B (ditto, Y)", we mean 
``A is smaller than B" and ``X is smaller than Y".

Given any function $f(x)$ defined for all $x\in A$, we define

\beq
\frac{f(x)}{\sum_{x\in A} numerator} =
\frac{f(x)}{\sum_{x\in A} f(x)} 
\;.
\eeq
This is just a shorthand, useful when
$f(x)$ is a long expression, 
to avoid writing $f(x)$ explicitly twice.

This paper will also utilize certain
notation associated with classical and quantum entropy.
See Refs.\cite{Tucci-review},\cite{Cover} for definitions and examples of  such notation.
In particular, we will assume that the reader is familiar with the definition of
the classical entropies $H(\rvx)$, $H(\rvx|\rvy)$ (conditional entropy) 
and $H(\rvx:\rvy)$ (mutual entropy) associated with 
any $P_\rvxy \in \pd(S_\rvxy)$. We will also assume that the reader is familiar
with the definitions of the quantum entropies 
$S_{\rho_\rvxy}(\rvx)$, $S_{\rho_\rvxy}(\rvx|\rvy)$ and $S_{\rho_\rvxy}(\rvx:\rvy)$
associated with any $\rho_\rvxy \in \dm(\hilxy)$.

For $P_\rvx, P'_\rvx \in \pd(S_\rvx)$, the {\it classical  
Kullback-Leibler (KL) distance}
is defined by

\beq
D(P_\rvx // P'_\rvx) = \sum_x P(x) \log_2 [P(x)/P'(x)]
\;.
\eeq
For $\rho_\rvx, \rho'_\rvx \in \dm(\hilx)$, the {\it quantum KL distance}
is defined by

\beq
D(\rho_\rvx // \rho'_\rvx) = \tr_\rvx [ \rho_\rvx( \log_2 \rho_\rvx - \log_2 \rho'_\rvx)]
\;.
\eeq
The classical (and quantum) KL distance is always non-negative and equals zero 
iff its two arguments are equal. It has many other useful properties. For more 
information about the KL distance, see
\cite{Cover} for the classical case and \cite{Wehrl} for the quantum one.

When discussing classical physics (ditto, quantum physics), we will refer
to various  probability distributions (ditto, density matrices) which are ``descendants" of
(i.e., can be derived from) a parent 
$P_\rvxyalp\in \pd(S_\rvxyalp)$ (ditto, $\alpxy{\rho}\in\dm(\hilxy)$).
As an aid to the reader, here is a table mapping the classical descendants  to their quantum counterparts.
The reader may find it helpful to continue returning to this ``Cast of Characters" table
as he advances through this play. 

\begin{center}
{\renewcommand{\arraystretch}{1.75}%change row height by factor
{\footnotesize
\begin{tabular}{|l|l|}\hline
 Classical & Quantum \\ 
\hline\hline
$ \{P(x, y| \alpha)|\forall x,y\} \in \pd(S_\rvxy) $&$ \alpxy{\rho} \in \dm(\hilxy) $\\
\hline
 $\{P(\alpha)|\forall \alpha\}\in \pd(S_\rvalp)$ & $\{ w_\alpha |\forall \alpha\} \in \pd(S_\rvalp)$\\
\hline
 $P(x, y, \alpha) = P(x, y|\alpha) P(\alpha)$ & $\alpxy{K} = w_\alpha \alpxy{\rho}$ \\                                             											  
  & $\rho_\rvxyalp = \sum_\alpha \proj{\alpha} \alpxy{K}$ \\                                             											  
\hline
 $P(x| \alpha) =\sum_y P(x, y| \alpha)$ & $\alpx{\rho} =\tr_\rvy \alpxy{\rho}$ \\                                             											  
\hline
 $P(x, \alpha) =\sum_y P(x, y, \alpha)$ & $\alpx{K} =\tr_\rvy \alpxy{K}$ \\                                             											  
\hline
 $P(y| \alpha) =\sum_x P(x, y| \alpha)$ & $\alpy{\rho} =\tr_\rvx \alpxy{\rho}$ \\                                             											  
\hline
 $P(y, \alpha) =\sum_x P(x, y, \alpha)$ & $\alpy{K} =\tr_\rvx \alpxy{K}$ \\                                             											  
\hline
 $P(x, y)=\sum_\alpha P(x, y, \alpha)$ & $\rho_\rvxy = \sum_\alpha \alpxy{K}$ \\                                             											  
\hline
 $R(x, y, \alpha) = \frac{P(x, \alpha)P(y, \alpha)}{P(\alpha)}$ & $\alpxy{R} = \frac{\alpx{K} \alpy{K}}{w_\alpha}$ \\                                             											  
\hline
\end{tabular}
}%footnotesize
}%stretch
\end{center}

In the classical case, we will use a functional $R[P_\rvxyalp]$ of $P_\rvxyalp$ defined
by

\beq
R[P_\rvxyalp](x, y, \alpha) =
\frac{P(x, \alpha) P(y, \alpha)}{P(\alpha)}
= P(x|\alpha)P(y|\alpha)P(\alpha)
\;.
\eeq
We will often abbreviate $R[P_\rvxyalp]$ by $R$ if no confusion is likely. Note that
$R(x, y, \alpha)\geq 0$ and $\sum_{x, y, \alpha}R(x, y, \alpha)=1$ so $R\in\pd(S_\rvxyalp)$.
In the quantum case, we will use a functional $\alpxy{R}[\alpxy{K}]$ of $\alpxy{K}$ defined
by

\beq
\alpxy{R}[\alpxy{K}] =
\frac{\alpx{K} \alpy{K}}{w_\alpha}
= w_\alpha \alpx{\rho}\alpy{\rho}
\;.
\eeq
We will often abbreviate $\alpxy{R}[\alpxy{K}]$ by $\alpxy{R}$ if no confusion is likely. Note that
$\alpxy{R}/w_\alpha\in\dm(\hilxy)$ for all $\alpha$.

Suppose $\rho_\rvxy\in\dm(\hilxy)$ has eigensystem 
$\{(\lambda_j , \ket{\phi_j}) | \forall j\}$. Thus, 

\beq
\rho_\rvxy = \sum_j \lambda_j | \phi_j \rangle\langle \phi_j |
\;.
\eeq
According to Ref.\cite{HJW}, 
$\rho_\rvxy$ can be expressed as

\beq
\rho_\rvxy = \sum_\alpha w_\alpha | \psi_\alpha \rangle\langle \psi_\alpha |
\;,
\eeq
where $w_\bullet\in \pd(S_\rvalp)$, and 
$\ket{\psi_\alpha}\in \hilxy$ for all $\alpha$,
if and only if there exists a transformation $T^\alpha_j$ 
($\alpha \in S_\rvalp$, $j\in S_\rvxy$)
which is ``right unitary":

\beq
\sum_\alpha T^\alpha_j T^{\alpha *}_{j'}= \delta^{j'}_j
\;,
\label{eq:right-uni}\eeq
and which satisfies

\beq
\sum_j T^\alpha_j \sqrt{\lambda_j} | \phi_j\rangle = 
\sqrt{w_\alpha} | \psi_\alpha\rangle
\;.
\eeq

Suppose $A:\hil\rarrow\hil$ is an operator with eigensystem 
$\{(\lambda_j , \ket{\phi_j}) | \forall j\}$. Thus, 

\beq
A= \sum_j \lambda_j \proj{\phi_j}
\;.
\eeq
The support (ditto, kernel) of $A$
is the subspace of $\hil$ consisting of the zero vector and all
those vectors in $\hil$ for 
which $A$ does not (ditto, does) vanish.
Suppose $\epsilon$ is a very small positive number and 
$\chi_{[0,\epsilon]}(x)$ for real $x$ is an indicator function that equals 1 if $x\in [0, \epsilon]$
and vanishes otherwise.
Then we define the projectors $\pi_{ker}(A)$ 
and $\pi_{supp}(A)$ by

\beq
\pi_{ker}(A) = \sum_j \chi_{[0,\epsilon]}(\lambda_j) \proj{\phi_j}
\;,
\eeq

\beq
\pi_{supp}(A) = 1 - \pi_{ker}(A)
\;.
\eeq

\section{Classical Physics Minimization}

In this section, we will discuss a  minimization of the CMI for classical probabilities.

The CMI for $P_\rvxyalp$  can be expressed as

\beqa
H(\rvx : \rvy | \rvalp)
& = & \sum_{x, y, \alpha} 
P(x, y, \alpha) \log_2 \left (
\frac{P(x, y | \alpha)}
{P(x | \alpha) P(y | \alpha) } 
\right ) 
\nonumber\\ 
& = & \sum_{x, y, \alpha} 
P(x, y, \alpha) \log_2 \left ( 
\frac{P(x, y, \alpha) P(\alpha)}
{P(x, \alpha) P(y, \alpha) } 
\right ) 
\nonumber\\ 
& = & \sum_{x, y, \alpha} 
P(x, y, \alpha) \log_2 \left ( 
\frac{P(x, y, \alpha)}
{R(x, y, \alpha) } 
\right ) 
\;.
\eeqa

Let ${\cal P}_{cla}$ be the set of those $P_\rvxyalp \in\pd(S_\rvxyalp)$ for which the sum over
$\alpha$ of $P(x, y, \alpha)$ equals a fixed $\tilde{P}_\rvxy \in \pd(S_\rvxy)$:

\beq
{\cal P}_{cla} = \{ P_\rvxyalp \in \pd(S_\rvxyalp) | P_\rvxy = \tilde{P}_\rvxy\}
\;.
\eeq
We define the entanglement $E_{cla}$ by 

\beq
E_{cla}=\left(\frac{1}{2}\right) \min_{P_\rvxyalp\in {\cal P}_{cla}} H(\rvx : \rvy | \rvalp) 
\;.
\eeq
In this definition of $E_{cla}$, $N_\rvalp$ is assumed to be fixed .
Clearly, for $N_{\rvalp}$ large enough, $E_{cla} = 0$. Indeed,
suppose $N_{\rvalp} = N_\rvx N_\rvy$ and 
$\alpha\rarrow (x_\alpha, y_\alpha)$ is a 1-1 onto function 
from $S_\rvalp$ to $S_{\rvx\rvy}$. 
Then $H(\rvx : \rvy | \rvalp)=0$
for 
$P(x, y, \alpha) = \tilde{P}(x_\alpha, y_\alpha) \delta_x^{x_\alpha} \delta_y^{ y_\alpha}$.

It is convenient to consider the following ``Lagrangian" functional of two probability distributions:

\beq
\lag(P_\rvxyalp, P'_\rvxyalp) = \sum_{x, y, \alpha} 
P(x, y, \alpha) \ln \left ( 
\frac{P(x, y, \alpha)}
{R'(x, y, \alpha)} 
\right ) 
\;,
\eeq
where $R' = R[P'_\rvxyalp]$.
$E_{cla}$ can be defined in terms of this Lagrangian by

\beq
E_{cla}= \left(\frac{1}{2 \ln 2}\right) \min_{P_\rvxyalp\in {\cal P}_{cla}}\lag ( P_\rvxyalp, P_\rvxyalp)
\;.
\eeq

\begin{lemma}
$\lag ( P_\rvxyalp, P'_\rvxyalp)$ is convex ($\cup$) in its first argument.
\end{lemma}
{\bf proof:}

The proof is
very similar to the proof that the 
classical entropy $H(P)$ is concave ($\cap$) in $P$.
Let $f(x) = x\ln x$. 
If we can show that 
 $\Delta\lag(P) = \sum_{x, y,\alpha} f(P(x, y, \alpha))$
is convex in $P$, we will be done, because the remaining part $\lag - \Delta\lag$
is linear in $P$.
Let $\lambda\in [0,1]$, $\bar{\lambda} = 1- \lambda$, and 
$P = \lambda P^{(1)} + \bar{\lambda} P^{(2)}$, where
$P^{(1)}, P^{(2)} \in pd(S_\rvxyalp)$. 
Since  $f(x)$ is convex in $x$, 

\beqa
\Delta\lag(P) 
&=& \sum_{x, y,\alpha} f\left(P(x, y, \alpha)\right) \nonumber\\
&\leq & \lambda
\sum_{x, y,\alpha} f\left( P^{(1)}(x, y, \alpha)\right)
+\bar{\lambda} 
\sum_{x, y,\alpha}f\left( P^{(2)}(x, y, \alpha)\right) \nonumber\\ 
&=& \lambda \Delta\lag( P^{(1)}) + \bar{\lambda} \Delta\lag( P^{(2)})
\;.
\eeqa
QED

(On the other hand, $\lag(P, P')$ is not generally convex or 
concave in its second argument. This can be seen
by taking the second derivative of $\lag$ with respect to that argument.)

\begin{theo}
Let

\beq
{\cal P}'_{cla} =\pd(S_\rvxyalp)
\;.
\eeq
At fixed $P_\rvxyalp\in {\cal P}_{cla}$,
$\lag(P_\rvxyalp, P'_\rvxyalp)$
is minimized over all $P'_\rvxyalp\in {\cal P}'_{cla}$
iff $P'_\rvxyalp$ satisfies

\begin{subequations} 
\beq
P'_\rvalp = P_\rvalp
\;,
\eeq

\beq
P'_{\rvx \rvalp} = P_{\rvx \rvalp}
\;,
\eeq
and

\beq
P'_{\rvy \rvalp} = P_{\rvy \rvalp}
\;.
\eeq
\label{eq:cla-p-prime-min}\end{subequations}
Thus,

\beq
\lag(P_\rvxyalp, P_\rvxyalp) = 
\min_{P'_\rvxyalp\in {\cal P}'_{cla}}
\lag(P_\rvxyalp, P'_\rvxyalp)
\;,
\eeq
and

\beq
E_{cla}= \left(\frac{1}{2 \ln 2}\right) 
\min_{P_\rvxyalp\in {\cal P}_{cla}}
\;\;
\min_{P'_\rvxyalp\in {\cal P}'_{cla}}
\lag ( P_\rvxyalp, P'_\rvxyalp)
\;.
\eeq
\end{theo}
{\bf proof:}

The Lagrangian $\lag$ 
can be expressed in terms of the KL distance as follows:

\beq
\frac{1}{\ln 2}\lag(P_\rvxyalp, P'_\rvxyalp) =
\left\{
\begin{array}{l}
D(P_\rvalp // P'_\rvalp) \\
+ \sum_\alpha P(\alpha) D\left(P(x, y|\alpha)//P(x|\alpha)P(y|\alpha) \right)\\
+ \sum_\alpha P(\alpha) D\left(P(x|\alpha) // P'(x|\alpha)\right) \\
+ \sum_\alpha P(\alpha) D\left(P(y|\alpha) // P'(y|\alpha)\right)
\end{array}
\right.
\;.
\eeq
The KL distance is always non-negative 
and equals zero iff its two arguments are equal. 
Hence, Eqs.(\ref{eq:cla-p-prime-min}) are necessary
and sufficient conditions for $\lag(P, P')$ to 
have a global minimum  in $P'$ at fixed $P$.
QED

\begin{theo}
At fixed $P'_\rvxyalp\in {\cal P'}_{cla}$,
$\lag (P_\rvxyalp, P'_\rvxyalp)$
is minimized over all $P_\rvxyalp\in {\cal P}_{cla}$
iff $P_\rvxyalp$ satisfies

\beq
P(\alpha | x, y) = R'(\alpha | x, y)
\;
\label{eq:cla-p-min}\eeq
for all $x, y , \alpha$.
\end{theo}
{\bf proof:}

Suppose a minimum is achieved.
Define a new Lagrangian $\lag_{tot}$ by adding to $\lag$ a Lagrange multiplier term
that enforces the constraint that the sum over $\alpha$ of $P(x, y, \alpha)$
equals a fixed probability distribution $\tilde{P}(x, y)\in \pd(S_\rvxy)$:

\beq
\lag_{tot} = \lag(P_\rvxyalp, P'_\rvxyalp)
+ \sum_{x,y} \lambda(x, y) [\sum_\alpha P(x, y, \alpha) - \tilde{P}(x, y)]
\;.
\eeq
$\lag_{tot}$ should not change if we vary infinitesimally and independently
the quantities $\lambda(x, y)$ and $P(x, y, \alpha)$
for all $x, y, \alpha$. Thus, 

\beq
0 = \pder{\lag_{tot}}{P(x_o, y_o, \alpha_o)} =
\ln  P(x_o, y_o, \alpha_o) + 1 -\ln R'(x_o, y_o, \alpha_o) + \lambda(x_o, y_o)
\;.
\eeq
Let

\beq
\Delta(x, y) = -1 - \lambda(x, y)
\;.
\eeq
Then

\beq
\ln P(x, y, \alpha) = \ln R'(x, y, \alpha) + \Delta(x, y)
\;.
\label{eq:cla-p-min-a}\eeq
Taking the exponential of both sides of the last equation and 
summing them over $\alpha$ yields the following constraint on $\Delta(x, y)$:

\beq
P(x, y) =  \sum_\alpha \exp[\ln R'(x, y, \alpha) + \Delta(x, y)]
\;.
\eeq
Solving the last equation for $\Delta(x, y)$ yields:

\beq
\Delta(x, y) = -\ln \left( 
\frac{R'(x,y)} {P(x,y)}\right )
\;.
\label{eq:cla-p-min-b}\eeq
Eq.(\ref{eq:cla-p-min}) now follows from Eqs.(\ref{eq:cla-p-min-a})
and (\ref{eq:cla-p-min-b}).

$\lag(P, P')$ has an extremum in $P$ at fixed $P'$ 
iff Eq.(\ref{eq:cla-p-min}) is true. 
Furthermore, since 
$\lag$ is convex in its first argument,
the extremum must be a global minimum.
QED

\begin{theo}
The CMI minimum which defines $E_{cla}$ is 
achieved iff

\beq
P(x, y, \alpha) = P(x, y) \frac{R(x, y, \alpha)}{\sum_\alpha R(x, y, \alpha)}
\;.
\label{eq:cla-final-conds}\eeq
Furthermore,

\beq
E_{cla}= \left(\frac{1}{2 \ln 2}\right) \av{\Delta}
\;,
\eeq
where

\beq
\Delta(x, y) = -\ln \left( 
\frac{R(x,y)} {P(x,y)}\right )
\;, 
\label{eq:cla-def-delta}\eeq
and

\beq
\av{\Delta} = \sum_{x, y} P(x, y) \Delta(x, y)
\;. 
\eeq
This justifies calling  $\Delta$ an ``entanglement operator".
\end{theo}
{\bf proof:}

The first part of this claim just
brings together results obtained in the previous two theorems.
The second part where $E_{cla}$ is expressed in terms of $\Delta$ follows from:

\beqa
E_{cla}&=& \left(\frac{1}{2 \ln 2}\right) \sum_{x, y, \alpha} P(x, y, \alpha) \ln \frac{P(x, y, \alpha)}{R(x, y, \alpha)}
\nonumber\\ 
&=& \left(\frac{1}{2 \ln 2}\right) \sum_{x, y, \alpha} P(x, y, \alpha)  \Delta(x, y)
\;.
\eeqa
QED

The last theorem gives certain conditions 
obeyed by any $P_\rvxyalp$ which achieves $E_{cla}$.
Next we will define a sequence of probability distributions, 
$P_\rvxyalp^{(n)}$ for $n=0, 1, \ldots$. The sequence will
converge to $P_\rvxyalp$ as $n\rarrow\infty$.
We will define our sequence recursively. 
In the following diagram, each quantity is defined in terms of
the quantities that point to it.

\beq
P_\rvxyalp^{(0)}\rarrow P_\rvxyalp^{(1)}\rarrow P_\rvxyalp^{(2)}\rarrow \cdots
\;
\eeq
Let $P_\rvxyalp^{(0)}$ be chosen arbitrarily from ${\cal P}_{cla}$.
For any $n\geq 0$, let

\beq
P^{(n+1)}(x, y, \alpha) =
P(x, y) 
\left[
\frac{
R^{(n)}(x, y, \alpha)
}{
\sum_{\alpha} R^{(n)}(x, y, \alpha)
}
\right]
\;,
\label{eq:cla-seq}\eeq
where $R^{(n)} = R[P^{(n)}_\rvxyalp]$.

In this paper, we won't prove that the sequence of 
$P_\rvxyalp^{(n)}$ converges. We defer that to future papers,
confining ourselves here to presenting
some empirical and  intuitive motivations for the sequence.
In Section \ref{sec:examples}, we give some 
computer results that are good empirical evidence of convergence.
Note that if the limit of the sequence does exist, then 
the limit of Eq.(\ref{eq:cla-seq}) is 
Eq.(\ref{eq:cla-final-conds}).

\section{Quantum Physics, Mixed Minimization}

In this section, we will discuss a quantum 
counterpart of the classical minimization problem discussed in the previous section.

Consider all $\rho_\rvxyalp\in \dm(\hilxyalp)$ of the special form:

\beq
\rho_\rvxyalp = \sum_\alpha  \proj{\alpha} w_\alpha\alpxy{\rho} = \sum_\alpha \proj{\alpha}\alpxy{K}
\;,
\eeq
where $\ket{\alpha}$ for all $\alpha$ is an orthonormal basis of $\hilalp$, 
$w_\bullet\in \pd(S_\alpha)$, and $\alpxy{\rho}\in \dm(\hilxy)$. 
As shown in Ref.\cite{Tucci-tang00b},
the CMI for $\rho_\rvxyalp$ can be expressed as

\beqa
S_{\rho_\rvxyalp}(\rvx: \rvy | \rvalp) 
&=& \sum_\alpha w_\alpha S_\alpxy{\rho} (\rvx : \rvy)
\nonumber\\ 
&=& \sum_\alpha w_\alpha [S(\alpx{\rho}) + S(\alpy{\rho}) - S(\alpxy{\rho})]
\nonumber\\ 
&=& \sum_\alpha \tr_{\rvx, \rvy}\left[ \alpxy{K} (\log_2 \alpxy{K} - \log_2 \frac{\alpx{K} \alpy{K}}{w_\alpha} )\right]
\nonumber\\ 
&=& \sum_\alpha \tr_{\rvx, \rvy}\left[ \alpxy{K} (\log_2 \alpxy{K} - \log_2 \alpxy{R} )\right]
\;.
\eeqa 

Let
 
\beq
{\cal K}_{mixed}^\star = \{ \dotxy{K} | \forall \alpha, 
\alpxy{K} = w_\alpha  \alpxy{\rho},
w_\bullet\in \pd(S_\rvalp), \alpxy{\rho}\in \dm(\hilxy) \}
\;,
\eeq
and

\beq
{\cal K}_{mixed}= \{ \dotxy{K} \in {\cal K}_{mixed}^\star| 
\sum_\alpha \alpxy{K} = \tilde{\rho}_{\rvx \rvy} \}
\;.
\eeq
We define the entanglement $E_{mixed}$ by

\beq
E_{mixed}= \left(\frac{1}{2}\right) \min_{ \dotxy{K} \in {\cal K}_{mixed}} S_{\rho_\rvxyalp}(\rvx: \rvy | \rvalp)
\;.
\eeq
In this definition of $E_{mixed}$,
$N_\rvalp$ will be assumed to tend to infinity.
 Ref.\cite{Horo-nsq} shows that
the limit is reached at a finite  $N_\rvalp \leq (N_\rvx N_\rvy)^2$.

It is convenient to consider the following ``Lagrangian" functional of two density matrices:

\beq
\lag(\alpxy{K}, \primealpxy{K}) = 
\sum_\alpha \tr_{\rvx, \rvy}\left[ \alpxy{K} (\ln \alpxy{K} - \ln \primealpxy{R} )\right]
\;,
\eeq
where $\primealpxy{R} = \alpxy{R}[\primealpxy{K}]$.
$E_{mixed}$ can be defined in terms of this Lagrangian by

\beq
E_{mixed}= \left(\frac{1}{2 \ln 2}\right) \min_{ \dotxy{K} \in {\cal K}_{mixed}} \lag(\alpxy{K}, \alpxy{K})
\;.
\eeq

\begin{lemma}
$\lag(\alpxy{K}, \primealpxy{K})$ is convex ($\cup$)
 in its first argument.
\end{lemma}
{\bf proof:}

The proof is very similar to the proof that the 
quantum entropy $S(\rho)$ is concave ($\cap$) in $\rho$
(see Ref.\cite{Wehrl}).
Let $f(x) = x\ln x$. 
If we can show that 
 $\Delta\lag(\alpxy{K}) = \sum_\alpha \tr_\rvxy f(\alpxy{K})$
is convex in $\alpxy{K}$, we will be done, because the remaining part $\lag - \Delta\lag$
is linear in $\alpxy{K}$.
Let $\lambda\in [0,1]$, $\bar{\lambda} = 1- \lambda$, and 
$K^{\alpha} = \lambda K^{(1)\alpha} + \bar{\lambda} K^{(2)\alpha}$.
Here $K^{(1)\alpha}$ and $K^{(2)\alpha}$ belong to $\dm(\hil_\rvxy)$
if we normalize them by
dividing them by their trace.
Let $K^{\alpha}$ have eigensystem $\{(m^\alpha_j, \ket{\phi^\alpha_j}) | \forall j\}$. 
Since  $f(x)$ is convex in $x$, 

\beqa
\Delta\lag(K^{\alpha}) 
&=& \sum_{\alpha, j} f(m^\alpha_j) = 
\sum_{\alpha, j}  f(\av{\phi^\alpha_j| K^\alpha | \phi^\alpha_j}) \nonumber\\
&\leq & \lambda
\sum_{\alpha, j} f(\av{\phi^\alpha_j| K^{(1)\alpha} | \phi^\alpha_j})
+\bar{\lambda} 
\sum_{\alpha, j} f(\av{\phi^\alpha_j| K^{(2)\alpha}  | \phi^\alpha_j}) \nonumber\\ 
&\leq & \lambda
\sum_{\alpha, j} \av{\phi^\alpha_j| f( K^{(1)\alpha}) | \phi^\alpha_j}
+\bar{\lambda} 
\sum_{\alpha, j} \av{\phi^\alpha_j| f(K^{(2)\alpha})  | \phi^\alpha_j} \nonumber\\ 
&=& \lambda \Delta\lag( K^{(1)\alpha}) + \bar{\lambda} \Delta\lag( K^{(2)\alpha})
\;.
\eeqa
QED

\begin{theo}
Let

\beq
{\cal K'}_{mixed}={\cal K}_{mixed}^\star
\;.
\eeq
At fixed $\dotxy{K} \in {\cal K}_{mixed}$,
$\lag(\alpxy{K}, \primealpxy{K})$
is minimized over all $\primedotxy{K} \in {\cal K'}_{mixed}$
iff $\primedotxy{K}$ satisfies

\begin{subequations} 
\beq
w'_\alpha = w_\alpha
\;,
\eeq

\beq
\primealpx{K} = \alpx{K}
\;,
\eeq
and

\beq
\primealpy{K} = \alpy{K}
\;.
\eeq
\label{eq:q-mixed-p-prime-min}\end{subequations}
Thus,

\beq
\lag(\alpxy{K}, \alpxy{K}) = 
\min_{ \primedotxy{K} \in {\cal K'}_{mixed}}
\lag(\alpxy{K}, \primealpxy{K})
\;,
\eeq
and

\beq
E_{mixed}= \left(\frac{1}{2 \ln 2}\right) 
\min_{ \dotxy{K} \in {\cal K}_{mixed}}
\;\;
\min_{ \primedotxy{K} \in {\cal K'}_{mixed}}
\lag (\alpxy{K}, \primealpxy{K})
\;.
\eeq
\end{theo}
{\bf proof:}

The Lagrangian $\lag$ 
can be expressed in terms of the KL distance as follows:

\beq
\frac{1}{\ln 2}\lag(\alpxy{K}, \primealpxy{K}) = 
\left\{
\begin{array}{l}
D(w_\alpha // w'_\alpha) \\
+ \sum_\alpha w_\alpha D(\alpxy{\rho}//\alpx{\rho}\alpy{\rho}) \\
+ \sum_\alpha w_\alpha D( \alpx{\rho} // \primealpx{\rho}) \\
+ \sum_\alpha w_\alpha D(\alpy{\rho} // \primealpy{\rho})
\end{array}
\right.
\;.
\eeq
Hence, Eqs.(\ref{eq:q-mixed-p-prime-min}) are necessary
and sufficient conditions for $\lag(K, K')$ to 
have a global minimum in $K'$ at fixed $K$.
QED

\begin{theo}
At fixed $\primedotxy{K} \in {\cal K'}_{mixed}$,
$\lag (\alpxy{K}, \primealpxy{K})$
is minimized over all $\dotxy{K} \in {\cal K}_{mixed}$
iff $\alpxy{K}$ satisfies

\begin{subequations}
\beq
\ln \alpxy{K} = \ln \primealpxy{R} + \Delta_\rvxy
\;,
\eeq
and

\beq
\rho_{\rvx \rvy} = \sum_\alpha \exp(\ln \primealpxy{R} + \Delta_\rvxy)
\;.
\eeq
\label{eq:q-mixed-p-min}\end{subequations}
\end{theo}
{\bf proof:}

Suppose a minimum is achieved.
Define a new Lagrangian $\lag_{tot}$ by adding to $\lag$ a Lagrange multiplier term
that enforces the constraint that the sum over $\alpha$ of $\alpxy{K}$
equals a fixed density matrix $\tilde{\rho}_\rvxy\in \dm(\hilxy)$:

\beq
\lag_{tot} = \lag(\alpxy{K}, \primealpxy{K})
+ \tr_{\rvx, \rvy} [\lambda_\rvxy (\sum_\alpha \alpxy{K} - \tilde{\rho}_\rvxy)]
\;.
\eeq
$\lag_{tot}$ should not change if we vary infinitesimally and independently the operators 
$\lambda_\rvxy$ and 
$\alpxy{K}$ 
for all $\alpha$. Thus, 

\beq
0=\delta \lag_{tot} = \sum_\alpha \tr_{\rvx, \rvy}\left [ \delta \alpxy{K} (
\ln \alpxy{K} +1 - \ln \primealpxy{R} + \lambda_\rvxy
)\right]
\;.
\eeq
Let

\beq
\Delta_\rvxy = -1 - \lambda_\rvxy
\;.
\eeq
Then

\beq
\ln \alpxy{K} = \ln \primealpxy{R} + \Delta_\rvxy
\;.
\label{eq:q-mixed-p-min-a}\eeq
Taking the exponential of both sides of the last equation and 
summing them over $\alpha$ yields the following constraint on $\Delta_\rvxy$:

\beq
\rho_\rvxy  = \sum_\alpha\exp[ \ln \primealpxy{R} + \Delta_\rvxy]
\;.
\label{eq:q-mixed-p-min-b}\eeq

$\lag(K, K')$ has an extremum in $K$ at fixed $K'$ 
iff Eqs.(\ref{eq:q-mixed-p-min}) are true. 
Furthermore, since 
$\lag$ is convex in its first argument,
the extremum must be a global minimum.
QED

\begin{theo}
The CMI minimum which defines $E_{mixed}$ is 
achieved iff

\begin{subequations}
\beq
\ln \alpxy{K} = \ln \alpxy{R} + \Delta_\rvxy
\;,
\label{eq:q-mixed-final-conds-k}\eeq
and

\beq
\rho_\rvxy  = \sum_\alpha \exp[\ln \alpxy{R} + \Delta_\rvxy]
\;.
\label{eq:q-mixed-final-conds-rho}\eeq
\label{eq:q-mixed-final-conds}\end{subequations}
Furthermore,

\beq
E_{mixed}= \left(\frac{1}{2 \ln 2}\right) \av{\Delta}
\;,
\eeq
where

\beq
\av{\Delta} = \tr_{\rvx, \rvy} (\rho_\rvxy \Delta_\rvxy)
\;.
\eeq
This justifies calling  $\Delta_\rvxy$ an ``entanglement operator".
\end{theo}
{\bf proof:}

The first part of this claim just
brings together results obtained in the previous two theorems.
The second part where $E_{mixed}$ is expressed in terms of $\Delta$ follows from:

\beqa
E_{mixed}&=& \left(\frac{1}{2 \ln 2}\right) \sum_\alpha \tr_{\rvx, \rvy}
[ \alpxy{K}( \ln\alpxy{K} - \ln\alpxy{R} ) ]
\nonumber\\ 
&=& \left(\frac{1}{2 \ln 2}\right) \tr_{\rvx, \rvy}(\rho_\rvxy \Delta_\rvxy)
\;.
\eeqa
QED

The last theorem gives certain conditions 
obeyed by any pair $(\alpxy{K},\Delta_\rvxy)$ which achieves $E_{mixed}$.
In the classical mixed minimization problem, 
we were able to solve for $\Delta$ explicitly and 
substitute it into the remaining equations. 
Non-commutativity now prevents us from doing this.
The way we will overcome the  obstacle 
of non-commutativity is to 
solve for both $\alpxy{K}$ and $\Delta_\rvxy$ simultaneously.
Next, we will define two sequences of operators,  
$K^{\alpha (n)}_\rvxy$ and $\Delta_\rvxy^{(n)}$ for $n=0, 1, 2\ldots$. The sequences will
converge to $\alpxy{K}$ and $\Delta_\rvxy$, respectively, as $n\rarrow\infty$.
We will define our two sequences recursively. 
In the following diagram, each quantity is defined in terms of
the quantities that point to it.\cite{alter}

\beq
\begin{array}{cccccc}
K^{\alpha (0)}_\rvxy&\rarrow &	K^{\alpha (1)}_\rvxy &\rarrow & K^{\alpha (2)}_\rvxy &\cdots \\
 &\nearrow&\downarrow &\nearrow & \downarrow & \\
\Delta^{(0)}_\rvxy	&\rarrow &\Delta^{(1)}_\rvxy &\rarrow &\Delta^{(2)}_\rvxy & \cdots
\end{array}
\;
\label{eq:k-jumps-first}\eeq

Roughly speaking, our strategy is: estimate $\alpxy{K}$, use the latter to get a better
estimate of $\Delta_\rvxy$, use the latter to get a better estimate of $\alpxy{K}$,
use the latter to get a better estimate of $\Delta_\rvxy$, and so on.
Let $K^{\alpha (0)}_\rvxy$ be chosen arbitrarily from ${\cal K}_{mixed}$.
Let $\Delta_\rvxy^{(0)}=0$.
For any $n\geq 0$, let

\begin{subequations}
\beq
K^{\alpha (n+1)}_\rvxy
=
\frac{
\pi_1
\exp \left[\ln R^{\alpha (n)}_\rvxy + \Delta_\rvxy^{(n)}\right]
\pi_1
}{
\sum_\alpha \tr_\rvxy (numerator)
}
\;,
\label{eq:q-mixed-seq-k}\eeq
(now use this $K$
to produce an even better $K$, which we call $K$ tilde:)

\beq
\tilde{K}^{\alpha (n+1)}_\rvxy
=
\frac{
\pi_1
\exp \left[\ln R^{\alpha (n+1)}_\rvxy + \Delta_\rvxy^{(n)}\right]
\pi_1
}{
\sum_\alpha \tr_\rvxy (numerator)
}
\;,
\label{eq:q-mixed-seq-ktilde}\eeq

\beq
I^{(n+1)} = 
\left( \pi_1\frac{1}{\sqrt{\rho_\rvxy}}\pi_1\right)
\left\{ 
\sum_\alpha
\tilde{K}^{\alpha (n+1)}_\rvxy
\right\}
\left( \pi_1\frac{1}{\sqrt{\rho_\rvxy}}\pi_1\right)
+\pi_0
\;,
\label{eq:q-mixed-seq-i}\eeq

\beq
\Delta_\rvxy^{(n+1)} =
-\ln 
\left(
e ^{\frac{-\Delta_\rvxy^{(n)}}{2}}
 I^{(n+1)}
 e ^{\frac{-\Delta_\rvxy^{(n)}}{2}}
 \right )
\;,
\label{eq:q-mixed-seq-delta}\eeq
\label{eq:q-mixed-seq}\end{subequations}
where $R^{\alpha (n)}_\rvxy = \alpxy{R}[K^{\alpha (n)}_\rvxy]$, 
$\pi_0 = \pi_{ker}(\rho_\rvxy)$ and $\pi_1 = 1 - \pi_0$.

In this paper, we won't prove that the sequences 
$K^{\alpha (n)}_\rvxy$ and $\Delta_\rvxy^{(n)}$ converge. 
We defer that to future papers,
confining ourselves here to presenting
some empirical and  intuitive motivations for the sequences.
In Section \ref{sec:examples}, we give some 
computer results that are good evidence of convergence.
It is easy to see that if the  sequences do converge,
then their limit satisfies Eqs.(\ref{eq:q-mixed-final-conds}). Indeed,
as $n \rarrow\infty$,
Eqs.(\ref{eq:q-mixed-seq}) become

\begin{subequations}
\beq
\alpxy{K}
=
\pi_1
\exp \left[\ln R^{\alpha }_\rvxy + \Delta_\rvxy\right]
\pi_1
\;.
\label{eq:q-mixed-seq-k-lim}\eeq

\beq
\rho_\rvxy = \sum_\alpha
\pi_1
\exp[
\ln R^{\alpha}_\rvxy
+ \Delta_\rvxy
]
\pi_1
\;,
\label{eq:q-mixed-seq-rho-lim}\eeq
\label{eq:q-mixed-seq-lim}\end{subequations}
Eq.(\ref{eq:q-mixed-seq-k-lim}) arises from combining the limits of
 Eqs.(\ref{eq:q-mixed-seq-k}) Eqs.(\ref{eq:q-mixed-seq-ktilde}).
Eq.(\ref{eq:q-mixed-seq-rho-lim}) arises from combining the limits of
Eqs.(\ref{eq:q-mixed-seq-i}) and (\ref{eq:q-mixed-seq-delta}).
Note also that when all operators are diagonal and therefore commute, 
Eqs.(\ref{eq:q-mixed-seq-i}) and (\ref{eq:q-mixed-seq-delta}) give
Eq.(\ref{eq:cla-def-delta}), the definition of the classical $\Delta$.
If we set $\pi_1 = 1$ and $\pi_0 = 0$ for now, then 
Eq.(\ref{eq:q-mixed-seq-k-lim}) is the same as 
Eq.(\ref{eq:q-mixed-final-conds-k}). And 
Eq.(\ref{eq:q-mixed-seq-rho-lim}) is the same as 
Eq.(\ref{eq:q-mixed-final-conds-rho}).

Now let us 
explain the purpose of the $\pi$ operators.
Ideally, we would want to define $I^{(n+1)}$ by

\beq
I^{(n+1)} = 
\frac{1}{\sqrt{\rho_\rvxy}}
\rho^{(n+1)}
\frac{1}{\sqrt{\rho_\rvxy}}
\;,
\label{eq:q-mixed-seq-naive-i}\eeq
where

\beq
\rho^{(n+1)} = 
\sum_\alpha
\tilde{K}^{\alpha (n+1)}_\rvxy
\;.
\eeq
However, if $\rho_\rvxy$ has any zero eigenvalues, its inverse square root does not exist.
Note $\rho^{(n+1)}$ tends to $\rho_\rvxy$. For some small positive real $\epsilon$, we can define

\beqa
\frac{1}{\sqrt{\rho_\rvxy}}
\rho^{(n+1)}
\frac{1}{\sqrt{\rho_\rvxy}}
&=& 
\frac{1}{\sqrt{\rho_\rvxy + \epsilon\pi_0}}
(\pi_1\rho^{(n+1)}\pi_1 + \epsilon\pi_0)
\frac{1}{\sqrt{\rho_\rvxy + \epsilon\pi_0}}\\
&=& 
\left(\pi_1\frac{1}{\sqrt{\rho_\rvxy}}\pi_1\right)
\rho^{(n+1)}
\left(\pi_1\frac{1}{\sqrt{\rho_\rvxy}}\pi_1\right)
+
\pi_0
\;.
\eeqa
(The $\epsilon\pi_0$ summands come into play only when $\rho_\rvxy$ vanishes.)
This 
justifies  Eq.(\ref{eq:q-mixed-seq-i}).
$\rho_\rvxy = \sum_\alpha \alpxy{K}$ so we must also require that
$\alpxy{K}$ vanish over the kernel space of $\rho_\rvxy$. We force this to happen 
by  pre and post
multiplying the right hand side of Eq.(\ref{eq:q-mixed-seq-k-lim}) by
$\pi_1$.

Another potential source of singular behavior in Eqs.(\ref{eq:q-mixed-seq}) is
the function $\exp(\ln R + D)$ where $R, D$ are Hermitian matrices
and $R$ can be singular.
This is not a theoretical disaster because even though the log of a zero eigenvalue of $R$
gives minus infinity, upon taking the exponential of that minus infinity, we get a zero contribution.
From a numerical point of view, calculating $\exp(\ln R + D)$ accurately
when $R$ is singular poses a challenge. Our first impulse is to calculate the 
eigenvalue expansion of $R$, take the log of the eigenvalues of the latter expansion, 
add $D$ to the result,
calculate the eigenvalue expansion of $\ln R + D$, exponentiate the eigenvalues of the latter expansion.
Finding the eigensystem of $\ln R + D$ can be hard to do accurately when $R$ is nearly singular
and therefore $\ln R + D$ contains some nearly infinite eigenvalues. 
There are, however, other ways of exponentiating a matrix which do not require 
calculating its eigensystem. Ref.\cite{Moller} describes 19 ``dubious" ways of 
exponentiating a matrix. Its authors use the adjective dubious because 
none of these methods is ideal. Some work only for certain types of matrices,
others entail an excessive number of operations, others are too sensitive, etc.
In our case, whenever we use $\exp(\ln R + D)$, the matrix
$\ln R + D$ is expected to be Hermitian with non-positive eigenvalues.
Method 4 of  
Ref.\cite{Moller} fits this situation perfectly. 
The method, first proposed by Colby et al in Ref.\cite{Cody}, is to 
approximate $\exp(-A)$
by a ratio of two $n$'th degree polynomials in A. 
The method works well even if
some of the eigenvalues of $A$ are nearly infinite, 
as long as they are all non-negative.

\section{Quantum Physics, Pure Minimization}

In this section, we will discuss another quantum 
 minimization problem.
This minimization will differ 
from the quantum mixed minimization discussed previously
in that now 
the range of our minimization will be restricted to those $\alpxy{\rho}\in \dm(\hilxy)$ of the special form:

\beq
\alpxy{\rho} = \proj{\psi_\alpha}
\;,
\eeq
where $\ket{\psi_\alpha}\in \hilxy$. 

As in the quantum mixed minimization problem, the CMI for $\rho_\rvxyalp$ can be expressed as

\beqa
S_{\rho_\rvxyalp}(\rvx: \rvy | \rvalp) 
&=& \sum_\alpha w_\alpha S_\alpxy{\rho} (\rvx : \rvy)
\nonumber\\ 
&=& \sum_\alpha w_\alpha [S(\alpx{\rho}) + S(\alpy{\rho}) - S(\alpxy{\rho})]
\nonumber\\ 
&=& \sum_\alpha \tr_{\rvx, \rvy}\left[ \alpxy{K} (\log_2 \alpxy{K} - \log_2 \alpxy{R} )\right]
\;.
\eeqa 

Let

\beq
{\cal K}_{pure}^\star = \{ \dotxy{K} | \forall \alpha, 
K^\alpha_\rvxy = w_\alpha  \proj{\psi_\alpha},
w_\bullet\in \pd(S_\rvalp), \ket{\psi_\alpha}\in \hilxy \}
\;,
\eeq
and

\beq
{\cal K}_{pure}= \{ \dotxy{K} \in {\cal K}_{pure}^\star| 
\sum_\alpha \alpxy{K} = \tilde{\rho}_{\rvx \rvy} \}
\;.
\eeq
We define the entanglement $E_{pure}$ by

\beq
E_{pure}= \left(\frac{1}{2}\right) \min_{ \dotxy{K} \in{\cal K}_{pure}} S_{\rho_\rvxyalp}(\rvx: \rvy | \rvalp)
\;.
\eeq
In this definition of $E_{pure}$,
$N_\rvalp$ will be assumed to tend to infinity.
 Ref.\cite{Horo-nsq} shows that
the limit is reached at a finite  $N_\rvalp \leq (N_\rvx N_\rvy)^2$.

Note that since we are now assuming that the $\alpxy{\rho}$ are pure for all $\alpha$,

\beqa
S_\alpxy{\rho} (\rvx : \rvy)
&=& S(\alpx{\rho}) + S(\alpy{\rho}) - S(\alpxy{\rho})
\nonumber\\ 
&=&2 S(\alpx{\rho})
\;.
\eeqa 
Note also that 

\beq
\alpx{\rho} = \tr_\rvy \proj{\psi_\alpha}
\;.
\eeq
If
$e = \{ (w_\alpha, \ket{\psi_\alpha}) | \alpha\in S_\rvalp\}$
where $\rho_\rvxy = \sum_\alpha w_\alpha \proj{\psi_\alpha}$, we call 
$e$ a $\rho_\rvxy$ {\it ensemble or preparation}. 
Thus, our definition of $E_{pure}$ can be re-expressed as

\beq
E_{pure} = \min_{e} \sum_\alpha w_\alpha S(\tr_\rvy \proj{\psi_\alpha}) 
\;,
\eeq
where the minimum is taken over  all $\rho_\rvxy$ ensembles  $e$. 
This is precisely the definition usually given for the entanglement of formation\cite{Ben}.
Thus, $E_{pure}$ is identical to the entanglement of formation.

It is convenient to consider the following ``Lagrangian" functional of two density matrices:

\beq
\lag(\alpxy{K}, \primealpxy{K}) = 
\sum_\alpha \tr_{\rvx, \rvy}\left[ \alpxy{K} (\ln \alpxy{K} - \ln \primealpxy{R} )\right]
\;,
\eeq
where $\primealpxy{R} = \alpxy{R}[\primealpxy{K}]$.
$E_{pure}$ can be defined in terms of this Lagrangian by

\beq
E_{pure}= \left(\frac{1}{2 \ln 2}\right) \min_{ \dotxy{K} \in {\cal K}_{pure}} \lag(\alpxy{K}, \alpxy{K})
\;.
\eeq

\begin{lemma}
$\lag(\alpxy{K}, \primealpxy{K})$ is convex ($\cup$)
 in its first argument.
\end{lemma}
{\bf proof:}

See proof of analogous lemma in the section on quantum mixed minimization.
QED

\begin{theo}
Let
\beq
{\cal K'}_{pure}={\cal K}_{pure}^\star
\;.
\eeq
At fixed $\dotxy{K} \in {\cal K}_{pure}$,
$\lag(\alpxy{K}, \primealpxy{K})$
is minimized over all $\primedotxy{K} \in {\cal K'}_{pure}$
iff $\primealpxy{K}$ satisfies

\begin{subequations}
\beq
w'_\alpha = w_\alpha
\;,
\eeq

\beq
\primealpx{K} = \alpx{K}
\;,
\eeq
and

\beq
\primealpy{K} = \alpy{K}
\;.
\eeq
\end{subequations}
Thus,

\beq
\lag(\alpxy{K}, \alpxy{K}) = 
\min_{ \primedotxy{K} \in {\cal K'}_{pure}}
\lag(\alpxy{K}, \primealpxy{K})
\;,
\eeq
and

\beq
E_{pure}= \left(\frac{1}{2 \ln 2}\right) 
\min_{ \dotxy{K} \in {\cal K}_{pure}}
\;\;
\min_{ \primedotxy{K} \in {\cal K'}_{pure}}
\lag (\alpxy{K}, \primealpxy{K})
\;.
\eeq
\end{theo}
{\bf proof:}

See proof of analogous theorem in the section on quantum mixed minimization.
QED

\begin{theo}
At fixed $\primedotxy{K} \in {\cal K'}_{pure}$,
$\lag (\alpxy{K}, \primealpxy{K})$
is minimized over all $\dotxy{K} \in {\cal K}_{pure}$
iff  $\alpxy{K} = w_\alpha \proj{\psi_\alpha}$ satisfies

\begin{subequations}
\beq
\ln(w_\alpha) \ket{\psi_\alpha} =
(\ln \primealpxy{R} + \Delta_\rvxy) \ket{\psi_\alpha}
\;,
\eeq
and

\beq
\rho_{\rvx \rvy} = \sum_\alpha w_\alpha \proj{\psi_\alpha}
\;.
\eeq
\label{eq:q-pure-p-min}\end{subequations}
\end{theo}
{\bf proof:}

Let
\beq
\ket{n_\alpha} = \sqrt{w_\alpha} \ket{\psi_\alpha}
\;
\eeq
and

\beq
A_\alpha = \ln \alpxy{K} - \ln \primealpxy{R}
\;.
\eeq
Then the Lagrangian $\lag$ can be expressed as:

\beq
\lag(\alpxy{K}, \primealpxy{K}) = \sum_\alpha \tr_\rvxy \left( \proj{n_\alpha} A_\alpha\right)
\;.
\eeq
Suppose a minimum is achieved.
Define a new Lagrangian $\lag_{tot}$ by adding to $\lag$ a Lagrange multiplier term
that enforces the constraint that the sum over $\alpha$ of $\proj{n_\alpha}$
equals a fixed density matrix $\tilde{\rho}_\rvxy\in \dm(\hilxy)$:

\beq
\lag_{tot} = \lag(\alpxy{K}, \primealpxy{K})
+ \tr_{\rvx, \rvy} [\lambda_\rvxy (\sum_\alpha \proj{n_\alpha} - \tilde{\rho}_\rvxy)]
\;.
\eeq
$\lag_{tot}$ should not change if we vary infinitesimally and independently the operators
$\lambda_\rvxy$ and 
 $\proj{n_\alpha}$ 
for all $\alpha$. Thus, 

\beq
0=\delta \lag_{tot} = \sum_\alpha \tr_{\rvx, \rvy}\left [ \delta ( \proj{n_\alpha}) (
A_\alpha +1  + \lambda_\rvxy
)\right]
\;.
\label{eq:q-pure-p-min-inter1}\eeq
Suppose we have an arbitrary Hermitian operator $A$ acting on some
Hilbert space $\hil$ and $\ket{n}\in \hil$. If

\beq
0=\tr \left [\delta (\proj{n}) A \right ]
\;,
\eeq
then

\beq
0=\bra{n} A \delta(\ket{n}) +
(\delta \bra{n}) A \ket{n}
\;
\eeq
so

\beq
A \ket{n} = 0
\;.
\eeq
Let

\beq
\Delta_\rvxy = -1 - \lambda_\rvxy
\;.
\eeq
Then Eq.(\ref{eq:q-pure-p-min-inter1}) implies that

\beq
(\ln \alpxy{K} - \ln \primealpxy{R} - \Delta_\rvxy)\ket{\psi_\alpha}=0
\;.
\label{eq:q-pure-p-min-inter2}\eeq
(Assume $w_\alpha\neq 0$ for all $\alpha$). Suppose $w\in(0,1]$. Given any column vector $\ket{\psi}\in\hil$, we can always find
a unitary matrix $U$ such that $\ket{\psi} = U\ket{0}$,
where $\ket{0}$ is the unit vector
which has one as its first component and zero for all others. Thus

\beqa
\ln \left(w\proj{\psi}\right) \ket{\psi} 
&=& U \ln \left(w \proj{0}\right) U^\dagger U \ket{0}
\nonumber\\ 
&=& U diag(\ln w, -\infty, -\infty, \ldots)
\left[ 
\begin{array}{c}
1\\0\\0\\ \vdots
\end{array}
\right]
\nonumber\\ 
&=& \ln(w) \ket{\psi}
\;.
\eeqa
Thus, Eq.(\ref{eq:q-pure-p-min-inter2}) implies that

\beq
\ln w_\alpha\ket{\psi_\alpha} =
(\ln \primealpxy{R} + \Delta_\rvxy)\ket{\psi_\alpha}
\;.
\eeq

$\lag(K, K')$ has an extremum in $K$ at fixed $K'$ 
iff Eqs.(\ref{eq:q-pure-p-min}) are true. 
Furthermore, since 
$\lag$ is convex in its first argument,
the extremum must be a global minimum.
QED

\begin{theo}
The CMI minimum which defines $E_{pure}$ is 
achieved  iff

\begin{subequations}
\beq
\ln(w_\alpha)\ket{\psi_\alpha} =
(\ln \alpxy{R} + \Delta_\rvxy) \ket{\psi_\alpha}
\;,
\label{eq:q-pure-final-conds-wpsi}\eeq
and

\beq
\rho_{\rvx \rvy} = \sum_\alpha w_\alpha \proj{\psi_\alpha}
\;.
\label{eq:q-pure-final-conds-rho}\eeq
\label{eq:q-pure-final-conds}\end{subequations}
Furthermore,

\beq
E_{pure}=\left(\frac{1}{2 \ln 2}\right) \av{\Delta}
\;,
\eeq
where

\beq
\av{\Delta} = \tr_{\rvx, \rvy} (\rho_\rvxy \Delta_\rvxy)
\;.
\eeq
This justifies calling  $\Delta_\rvxy$ an ``entanglement operator".
\end{theo}
{\bf proof:}

The first part of this claim just
brings together results obtained in the previous two theorems.
For a proof of the second part where $E_{pure}$ is expressed in terms of $\Delta$, 
see the analogous theorem for quantum mixed minimization.
QED

As in the quantum mixed minimization problem, 
we will define two sequences of operators,  
$K^{\alpha (n)}_\rvxy = w_\alpha^{(n)}\proj{\psi_\alpha^{(n)}}$ 
and $\Delta_\rvxy^{(n)}$ for $n=0, 1, 2\ldots$. The sequences will
converge to $\alpxy{K}$ and $\Delta_\rvxy$, respectively, as $n\rarrow\infty$.
We will define our two sequences recursively. 
Let $K^{\alpha (0)}_\rvxy $ be chosen arbitrarily from ${\cal K}_{pure}$.
Let $\Delta_\rvxy^{(0)}=0$.
For any $n\geq 0$, let

\begin{subequations}
\beq
w_\alpha^{(n+1)}\ket{\psi_\alpha^{(n+1)}} =
\pi_1
\exp\left[ 
\ln R^{\alpha (n)}_\rvxy
+ \Delta^{(n)}_\rvxy
\right]
\pi_1 
\ket{\psi_\alpha^{(n+1)}}
\;,
\label{eq:q-pure-seq-wpsi}\eeq

\beq
K^{\alpha (n+1)}_\rvxy
=
\frac{
w_\alpha^{(n+1)} \proj{\psi_\alpha^{(n+1)}}
}{
\sum_\alpha \tr_\rvxy (numerator)
}
\;,
\label{eq:q-pure-seq-k}\eeq

\beq
\tilde{w}_\alpha^{(n+1)}\ket{\tilde{\psi}_\alpha^{(n+1)}} =
\pi_1
\exp\left[ 
\ln R^{\alpha (n+1)}_\rvxy
+ \Delta^{(n)}_\rvxy
\right]
\pi_1 
\ket{\tilde{\psi}_\alpha^{(n+1)}}
\;,
\label{eq:q-pure-seq-wpsi-tilde}\eeq

\beq
\tilde{K}^{\alpha (n+1)}_\rvxy
=
\frac{
\tilde{w}_\alpha^{(n+1)} \proj{\tilde{\psi}_\alpha^{(n+1)}}
}{
\sum_\alpha \tr_\rvxy (numerator)
}
\;,
\label{eq:q-pure-seq-k-tilde}\eeq

\beq
I^{(n+1)} = 
\left( \pi_1\frac{1}{\sqrt{\rho_\rvxy}}\pi_1\right)
\left\{ 
\sum_\alpha
\tilde{K}^{\alpha (n+1)}_\rvxy
\right\}
\left( \pi_1\frac{1}{\sqrt{\rho_\rvxy}}\pi_1\right)
+\pi_0
\;,
\label{eq:q-pure-seq-i}\eeq

\beq
\Delta_\rvxy^{(n+1)} =
-\ln 
\left(
e ^{\frac{-\Delta_\rvxy^{(n)}}{2}}
 I^{(n+1)}
 e ^{\frac{-\Delta_\rvxy^{(n)}}{2}}
 \right )
\;,
\label{eq:q-pure-seq-delta}\eeq
\label{eq:q-pure-seq}\end{subequations}
where $R^{\alpha (n)}_\rvxy = \alpxy{R}[K^{\alpha (n)}_\rvxy]$, 
$\pi_0 = \pi_{ker}(\rho_\rvxy)$ and $\pi_1 = 1 - \pi_0$.
In Eqs.(\ref{eq:q-pure-seq-wpsi}) and (\ref{eq:q-pure-seq-wpsi-tilde}),
 we choose the {\bf largest} eigenvalue and the corresponding eigenvector
of the operator on the right hand side of the equation. As $n$ tends to infinity,
said operator tends towards a projection operator, so its
eigenvalues all go to zero except for possibly one of them. 

In this paper, we won't prove that the sequences 
$K^{\alpha (n)}_\rvxy$ and $\Delta_\rvxy^{(n)}$ converge. 
We defer that to future papers,
confining ourselves here to presenting
some empirical and  intuitive motivations for the sequences.
In Section \ref{sec:examples}, we give some 
computer results that are good evidence of convergence.
It is easy to see that if the  sequences do converge,
then their limit satisfies Eqs.(\ref{eq:q-pure-final-conds}).

\section{Causa Com\'{u}n}\label{sec:examples}
We have written a C++ program called Causa Com\'{u}n
(``Common Cause" in Spanish). It's called this
because entanglement is a manifestation of causality; it
occurs between several events with a common cause.
Causa Com\'{u}n can do all three 
minimizations considered in this paper (classical, quantum mixed and quantum pure).
For each of these three cases, it can find the entanglement, 
entanglement operator, and an optimal state decomposition.
Next, we will discuss Causa Com\'{u}n output for two examples of quantum states:
Bell Mixtures, and Horodecki States.

\subsection{Bell Mixtures}

In this example 

\beq
S_\rvx = S_\rvy = Bool
\;.
\eeq
The following four states are usually called the ``Bell basis" of $\hilxy$:
\beq
\ket{\psi^\pm} = \ket{\neq^\pm} = 
\frac{1}{\sqrt{2}}( \ket{01} \pm \ket{10} )
\;,
\eeq
and

\beq
\ket{\phi^\pm} = \ket{=^\pm} = 
\frac{1}{\sqrt{2}}( \ket{00} \pm \ket{11} )
\;.
\eeq

Let
$\ket{B(\mu)}$
with $\mu \in \{0,1,2,3\}$ represent the four Bell states.
Call a ``Bell mixture" any density matrix $\rho_\rvxy$
expressible in the form

\beq
\rho_\rvxy = \sum_\mu m_\mu \ket{B(\mu)}\bra{B(\mu)}
\;,
\eeq
where $m_\bullet \in \pd(S_\rvmu)$.
For any $p\in [0, 1]$, define the {\it binary entropy} $h(p)$  by 

\beq
h(p) = -[p \log_2 (p)+ (1-p) \log_2 (1-p)]
\;.
\eeq
Ref.\cite{Ben} showed that for any Bell mixture $\rho_\rvxy$,
the entanglement of formation is given by:

\begin{subequations}
\beq
E_{form.} (\rho_\rvxy) = 
h\left( \frac{1+\sqrt{1-t}}{2} \right)
\;,
\eeq

\beq
t = 
\left \{
\begin{array}{ll}
0& {\rm if}\; m_{max} < \frac{1}{2}\\
(2m_{max} -1)^2 & {\rm otherwise}
\end{array}
\right.
\;,
\eeq
\label{eq:woo-formula}\end{subequations}
where $m_{max}$ refers to the maximum of the weights $m_\mu$.

We will refer to the following one parameter family of Bell mixtures
as the ``Werner States":

\beq
W(F) = 
F\ket{=^+}\bra{=^+} 
+ \frac{(1-F)}{3} \left(
\ket{=^-}\bra{=^-} 
+ \ket{\neq^+}\bra{\neq^+} 
+ \ket{\neq^-}\bra{\neq^-} 
\right)
\;,
\eeq
for $F\in[0,1]$.

\begin{figure}[h]
	\begin{center}
	\epsfig{file=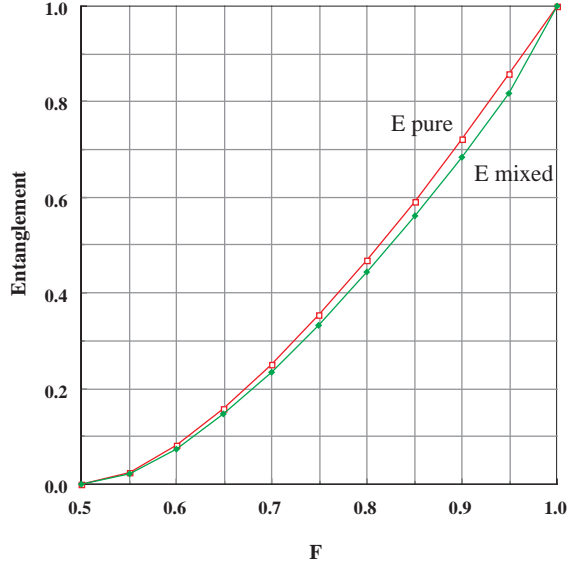, height=3.0in}
	\caption{Pure and mixed entanglements for Werner States.}
	\label{fig:werner}
	\end{center}
\end{figure}

Fig.\ref{fig:werner} is Causa Com{\'u}n output for
$E_{pure}$ and $E_{mixed}$ of 
the Werner States using $N_\rvalp = 6$. We also got $E_{mixed} = E_{pure} = 0$
for $0\leq F\leq \frac{1}{2}$ (not shown in graph). We compared
$E_{pure}$ obtained using our algorithm
and $E_{form}$ obtained using Eq.(\ref{eq:woo-formula}), and found them to
agree well over the entire range $F\in[0,1]$.
Note that $E_{mixed}$ behaves like the entanglement of distillation:
both are non-negative, less than or equal to $E_{form}$, 
and equal to $E_{form}$ for pure density matrices. In a future paper,
we will explain the close connection between $E_{mixed}$ and entanglement of 
distillation.

\subsection{Horodecki States}

In this example, 

\beq
S_\rvx = S_\rvy = \{0,1,2\}
\;.
\eeq
Let

\beq
\av{x,y | \psi^+} =
\frac{1}{\sqrt{3}}
(\delta_{xy}^{00} + \delta_{xy}^{11} + \delta_{xy}^{22})
\;,
\eeq

\beq
(\sigma^+_\rvxy)_{xy, x'y'} = 
\frac{\delta_{xy}^{x'y'}}{3}
(\delta_{xy}^{01} + \delta_{xy}^{12} + \delta_{xy}^{20})
\;,
\eeq

\beq
(\sigma^-_\rvxy)_{xy, x'y'} = 
\frac{\delta_{xy}^{x'y'}}{3}
(\delta_{xy}^{10} + \delta_{xy}^{21} + \delta_{xy}^{02})
\;.
\eeq
We will refer to the following one parameter family of density matrices
as the ``Horodecki States":

\beq
\sigma(\alpha) = 
\frac{2}{7} \ket{\psi^+}\bra{\psi^+}
+\frac{\alpha}{7} \sigma^+
+\frac{5-\alpha}{7} \sigma^-
\;,
\eeq
where $\alpha\in [2,5]$.
These states where first introduced in Ref.(\cite{Horo-b-ent}),
where it was shown that they are separable for $\alpha\in(2,3)$,
bound entangled for $\alpha\in(3,4)$, 
and free entangled for $\alpha\in(4,5)$.

\begin{figure}[h]
	\begin{center}
	\epsfig{file=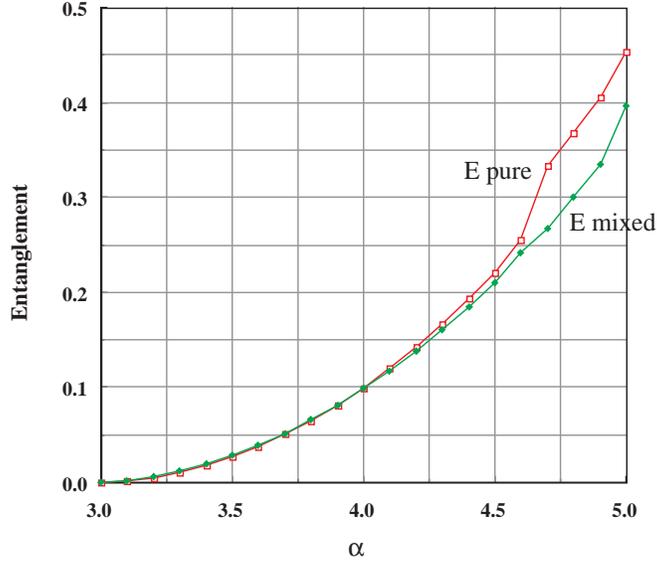, height=3.0in}
	\caption{Pure and mixed entanglements for Horodecki States.}
	\label{fig:horo}
	\end{center}
\end{figure}

Fig.\ref{fig:horo} is Causa Com{\'u}n output for
$E_{pure}$ and $E_{mixed}$ of 
the Horodecki States using $N_\rvalp = 12$.
We also got $E_{mixed} = E_{pure} = 0$
for $2\leq \alpha\leq 3$ (not shown in graph).\cite{startingpoints}

%\section{Nomenclature}
%\section{Conclusions}

\section*{Acknowledgements}
I thank K. Audenaert and K. Zyczkowski for interesting email 
about entanglement.


\begin{thebibliography}{99}

\bibitem{Tucci-tang99}
R.R. Tucci,
``Quantum Entanglement and Conditional Information Transmission", 
Los Alamos eprint quant-ph/9909040 .

\bibitem{Tucci-tang00a}
R.R. Tucci,
``Separability of Density Matrices and
Conditional Information Transmission", 
Los Alamos eprint quant-ph/0005119 .

\bibitem{Tucci-tang00b}
R.R. Tucci,
``Entanglement of formation and conditional information transmission",
Los Alamos eprint quant-ph/0010041 .

\bibitem{Ari} 
S. Arimoto,
``An algorithm for calculating the capacity of an
arbitrary discrete memoryless channel", 
IEEE Trans. Info Theory {\bf 18}, 14 (1972).

\bibitem{Bla}
R. E. Blahut,
``Computation of channel capacity and rate distortion functions", 
IEEE Trans. Info Theory {\bf 18}, 460 (1972).

\bibitem{Bla-book}
R.E. Blahut, ``Principles and Practice of Information Theory",
(Addison-Wesley, 1987) pgs 164-170.

\bibitem{Osawa} S. Osawa, H. Nagaoka,
``Numerical experiments on the capacity of a quantum channel with entangled inputs",
Los Alamos eprint quant-ph/0007115 .

\bibitem{Zyc}
K. Zyczkowski,
``Volume of the set of entangled states II",
Los Alamos eprint quant-ph/9902050 .

\bibitem{dutch-boys}
K. Audenaert, F. Verstraete, B. De Moor,
``Variational characterizations of separability and entanglement of formation",
Los Alamos eprint quant-ph/0006128 .


\bibitem{Tucci-review}
R.R. Tucci,
``Quantum Information Theory - A Quantum Bayesian Nets Perspective", 
Los Alamos eprint quant-ph/9909039 .

\bibitem{Cover}
Thomas M. Cover and Joy A. Thomas, ``Elements of Information Theory (2cnd edition)"
(Wiley, 1991).

\bibitem{Wehrl}
A. Wehrl, ``General Properties of Entropy",
Rev. Mod. Phys. {\bf 50} 221-260 (1978).

\bibitem{alter}Call diagram Eq.(\ref{eq:k-jumps-first}) a K-jumps-first strategy. It is also possible to use a
$\Delta$-jumps-first strategy defined by the following diagram:

\beq
\begin{array}{cccccc}
K^{\alpha (0)}_\rvxy&\rarrow &	K^{\alpha (1)}_\rvxy &\rarrow & K^{\alpha (2)}_\rvxy &\cdots \\
 &\searrow&\uparrow &\searrow & \uparrow & \\
\Delta^{(0)}_\rvxy	&\rarrow &\Delta^{(1)}_\rvxy &\rarrow &\Delta^{(2)}_\rvxy & \cdots
\end{array}
\;
\label{eq:delta-jumps-first}\eeq
We won't discuss the $\Delta$-jumps-first strategy in this paper because it is very similar to the
$K$-jumps-first one. (The $\Delta$-jumps-first strategy appears 
to require more numerical operations to move from $n$ to $n+1$)

\bibitem{Horo-nsq} 
P. Horodecki,
``Separability Criterion and Inseparable Mixed States
with Positive Partial Transpose",
Los Alamos eprint quant-ph/9703004 .

\bibitem{Moller}
C.B. Moller, C.F. Van Loan,
``19 dubious ways to compute the exponential of a matrix",
SIAM Review {\bf 20}, 801 (1979).

\bibitem{Cody}
W.J. Cody, G. Meinardus, R.S. Varga,
``Chebyshev Rational Approximations to $e^{-x}$ in $[0, +\infty)$
and applications to heat conduction problems", 
Jour. of Approx. Theo. {\bf 2}, 50 (1969).


\bibitem{HJW}
L.P. Hughston, R. Jozsa, W.K. Wootters,
``A complete classification of quantum ensembles having a given density matrix",
Phys. Let. A {\bf 183} (1993) 14.


\bibitem{Ben}
C.H. Bennett, D.P. DiVincenzo, J. Smolin, W.K. Wootters,
``Mixed state entanglement and quantum error correction",
Los Alamos eprint quant-ph/9604024 .


\bibitem{Horo-b-ent} 
P. Horodecki,
``Bound entanglement can be activated",
Los Alamos eprint quant-ph/9806058 .

\bibitem{startingpoints} 
K. Audenaert (KA)  et al have written a  computer program that calculates entanglement of formation
using a method that is very different from mine. KA kindly provided me with a plot 
generated by his program
of $E_{form}$ versus $\alpha$ for the Horodecki States.
I compared his $E_{form}$ with my $E_{pure}$.
I found both graphs to agree very well
for $\alpha\in[2,4.5]$ and for $\alpha=5$. His graph
connects the points at $\alpha=4.5$ and $\alpha=5$ in a smooth way.
Mine is higher in that range although we agree at 4.5 and 5. 
I believe his plot to be the more accurate.
I believe the source of our disagreement is that Causa  Com\'{u}n
is getting stuck in non-global minima.

KA et al  reported in Ref.(\cite{dutch-boys}) that they
routinely run their program with about ten starting points and 
then choose the smallest $E_{form}$ of the ten. They do this because they
have found  multiple minima in the function being minimized.
I too average over several (five) starting points as I too have
found multiple minima. 

I am currently using poorly motivated starting points:
I generate a random 
right unitary matrix $T^\alpha_j$ 
and construct $K^{\alpha(0)}_\rvxy$ from $T^\alpha_j$ and $\rho_\rvxy$.
KA et al, on the other hand, use a technique (see Ref.(\cite{dutch-boys}) which gives them a 
more refined guess as their starting point. I intend to incorporate 
a similar technique into Causa  Com\'{u}n in the future. I 
believe that by choosing a starting point $K^{\alpha(0)}_\rvxy$ with the appropriate symmetry,
one can significantly improve the chances that the algorithm will converge to the global minimum.
An analogous issue arises when using a variational method to 
solve for the quantum stationary states of a particle in a potential well.
Using a trial wavefunction with the appropriate symmetry 
(e.g., no nodes)
helps the algorithm converge to the ground state energy level.

\end{thebibliography}
\end{document}